\documentclass[aps,prl,twocolumn,groupedaddress,showpacs]{revtex4}

\usepackage{graphicx}

\begin{document}

\title{Radio-frequency discharges in Oxygen. Part 1: Modeling}

\author{F. X. Bronold$^1$, K. Matyash$^2$, D. Tskhakaya$^{3}$, R. Schneider$^2$ and H. Fehske$^1$}
\address{$^1$Institut f\"ur Physik, Ernst-Moritz-Arndt-Universit\"at Greifswald, D-17489
Greifswald, Germany}
\affiliation{$^2$Max-Planck-Institut f\"ur Plasmaphysik, Teilinstitut Greifswald, D-17491
Greifswald, Germany}
\affiliation{$^3$Institut f\"ur Theoretische Physik, Universit\"at Innsbruck, A-6020
Innsbruck, Austria}

\date{\today}

\begin{abstract}
In this series of three papers we present results from a combined experimental 
and theoretical effort to quantitatively describe capacitively coupled radio-frequency 
discharges in oxygen. The particle-in-cell Monte-Carlo model on which the theoretical 
description is based will be described in the present paper. It treats space charge 
fields and transport processes on an equal footing with the most important 
plasma-chemical reactions. For given external voltage and pressure, the model  
determines the electric potential within the discharge and the distribution functions 
for electrons, negatively charged atomic oxygen, and positively charged molecular oxygen.
Previously used scattering and reaction cross section data are critically 
assessed and in some cases modified. To validate our model, we compare the densities 
in the bulk of the discharge with experimental data and find good agreement, indicating 
that essential aspects of an oxygen discharge are captured.  
\end{abstract}

\pacs{52., 52.35.Tc, 82.33.Xj, 47.11.-j}
\maketitle

\section{\label{Intro}I. Introduction}
Electro-negative gases, such as ${\rm SF_6}$, ${\rm Cl_2}$, ${\rm CF_4}$, and 
${\rm O_2}$, play an important role in plasma-assisted materials processing. 
Most notably ${\rm O}_2$, either as the main feedstock gas or as an admixture 
to halogen- or silicon-based gases, is of vital importance for a large 
variety of etching and thin-film deposition 
techniques~\cite{Tolliver84,SP89,CHL91,KKS92,KJ96,NGG98,MZK98}. The requirements 
on the controllability and predictability of these processes are so high that 
further advancement of this technology will depend on modeling tools which 
go beyond the macroscopic, fluid-type approximations, which, for
instance, cannot reliably predict the velocity distributions of the species.

Although electro-negative gas discharges have been studied for a long 
time~\cite{Seeliger49,Sabadil73,EE80}, with significant progress made during the last 
two decades~\cite{FGT88,Tsendin89,LVL94,FS99}, a complete quantitative description 
is still lacking, in particular, with respect to discharge profiles, structuring, and 
operation regimes~\cite{Franklin01b,Franklin02}. This is  not surprising because
the occurrence of negative ions leads to abrupt changes in the ion density 
(density fronts~\cite{Kaganovich01}) which in most cases force the discharge to 
stratify into a central quasi-neutral ion-ion plasma and a peripheral electro-positive
electron-ion plasma~\cite{EE80,FGT88,Tsendin89,LVL94,FS99,FS00,LMF04}. 
The transition between the two is rather subtle. It can be, for instance,  
accompanied by a double layer (internal sheath~\cite{KLL99}). 
Electro-negative gas discharges are thus rather complex and the investigation of  
the spatio-temporal structure of the discharge as a function of external control 
parameters (current, voltage, frequency, pressure, and geometry) is 
a great theoretical~\cite{Boeuf87,Shveigert91,Kaganovich95,LIL06} and 
experimental~\cite{QDG98,IHS98,KSQ00,BSB00,Kono02,Dittmann07} challenge. In 
addition, electro-negative processing gases are reactive molecular gases, with 
internal degrees of freedom, which lead to a quite involved plasma chemistry.

Similar to the kinetic description of a mixture of reacting gases~\cite{BR90}, the 
modeling of a gas discharge could be based on a coupled set of Boltzmann equations 
for the distribution functions of the species which need to be treated kinetically. 
This set has to be augmented by Maxwell's equations, or parts of it, depending on 
how the discharge is electrically driven. Even for simple reactive gas discharges, 
with only two negatively and one positively charged species, this 
approach is not practical. More promising are methods which track representative 
samples of simulated particles subject to (non-reactive and reactive) collisions 
as well as electromagnetic fields, which are again determined from the relevant 
parts of Maxwell's equations. Although these approaches are closely related to 
the direct simulation method successfully employed for the modeling of the flow 
of rarefied gases~\cite{Bird69,Nanbu80,Bird94}, in the context of gas 
discharges~\cite{Nanbu00}, it is more common to refer to them as particle-in-cell 
Monte Carlo collision (PIC-MCC) methods~\cite{VS95,KBV00,TK02,Verboncoeur05}, because 
without collisions the methods collapse to the PIC approach~\cite{Birdsall85,Hockney89} 
for the solution of the Vlasov problem.   

This is the first paper in a series of three where we report the results 
of a combined experimental and theoretical study of the interplay 
between plasma-chemistry and electrodynamics in capacitively coupled
radio-frequency (rf) discharges in oxygen. 
Oxygen is a weakly electro-negative gas whose plasma chemistry is strongly 
affected by vibrational, rotational and meta-stable states. Up to 75 reaction and 
scattering processes have been listed to potentially affect the properties of the 
discharge~\cite{KG03}. Not all of them can be equally important. The challenge is 
therefore to identify (via comparison between experiment and modeling) the subset 
of collisions responsible for the experimental findings under consideration. 

In the present paper, we describe such a reduced model. It is tailor-made for the
investigation of the charged species of the discharge, including the formation
of ion density fronts, internal sheaths, and the resulting stratification 
of the discharge. We critically assess the cross sections used to characterize
the selected collision channels and point out some inconsistencies in previously 
used cross section data. The validity of our model is verified by a comparison of 
calculated bulk densities with experimentally measured ones~\cite{KSQ00}. 
The following two papers discuss, from the experimental~\cite{Dittmann07} and 
theoretical~\cite{Matyash07} point of view, respectively, the sheath region of the 
discharge focusing on the $844~nm$ double emission layer in front of 
the powered electrode and a comparison of experimental and simulated distribution 
functions for positive ions.

The next section describes the main features of our PIC-MCC
implementation~\cite{TA77,Matyash03}. It is an extension of a code, which was 
originally designed for the particle-based modeling of sheaths in magnetized
plasmas~\cite{Chodura82,Bergmann94}, to the case of electrically driven 
discharges in reactive molecular gases. We focus, in particular, on the handling 
of collision processes, which deviates from the treatments usually employed in 
the plasma context~\cite{VS95,TK02,Verboncoeur05} and is closer to the 
simulation approaches used in rarefied gas dynamics~\cite{Bird69,Bird94}. The 
collisions included in our model are discussed in Section III. Where necessary, 
we combine measured cross 
sections~\cite{Phelps85,GR72,STS63,Muschlitz60,RB65,TB68,PFD79,VKM96,WD74,PP98,CS74a,CS74b,BM70}
with model cross sections~\cite{Olson72,OST78,Bardsley68,AKO82,Lieberman05}
to characterize collisions and fix the free parameters through a comparison 
with experimental data. In Section IV we show that once the parameters are 
fixed, good agreement between simulation and experiment can be achieved in 
the bulk of the discharge. 
Section V summarizes the essentials of our model and concludes with a short outlook.  

\section{II. Method of simulation}

\begin{figure}
\begin{center}
\includegraphics[width=0.7\linewidth]{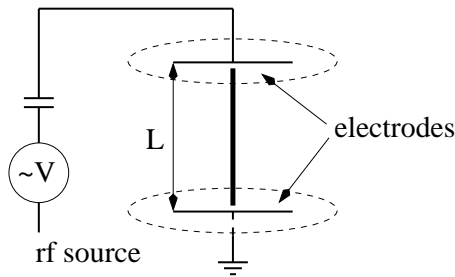}
\caption{\label{geometry}
Schematic geometry of the rf discharges used in Refs.~\cite{KSQ00,Dittmann07}.
We use a planar, one-dimensional model to simulate the central axial part of the
discharge (thick solid line).
}
\end{center}
\end{figure}
We are only interested in the central axial part of the rf discharges
described in Refs.~\cite{KSQ00,Dittmann07}. Ignoring the electric
asymmetry between the powered and grounded electrode, 
this part of the discharge can be simulated by the planar, one-dimensional 
model shown in figure~\ref{geometry}. The planar model retains only one 
spatial coordinate $x$, with $0\le x\le L$, where $L$ is the distance between
the electrodes, but it keeps the full three-dimensional velocity space for the 
particles (1d3v model). One of the electrodes is driven by the rf voltage 
$U(t)=U_{rf}\sin(2\pi f_{rf}t)$, with $f_{rf}=13.6~MHz$ and $U_{rf}$ ranging 
from $75~V$ to $800~V$, while the other is grounded. Both electrodes are assumed to 
be totally absorbing; secondary electron emission is neglected. To mimic the constant 
oxygen flow through the discharge chamber, we enforce in the simulation volume a 
constant oxygen pressure ranging from $10~Pa$ to $100~Pa$.

The simulation volume of the planar model is of course not well defined because the 
lateral ``cross-section'' A is a free parameter. This parameter actually controls the 
weight of the simulated particles, that is, the number of real particles represented 
by one simulated particle. 
We arbitrarily choose $A=\lambda_{De}^2$ with $\lambda_{De}$ the 
electron Debye length of a reference electron system (RES) which we use to 
initially set-up
the length and time scales of the simulation (see below). The RES is specified 
by a density $n_{res}$ and a temperature $T_{res}$. Both have to be
adjusted to the particular experimental conditions. For the  
experiments~\cite{KSQ00,Dittmann07}, 
$n_{res}\approx 10^8~$--$~10^9~cm^{-3}$ and $k_BT_{res}~\approx~10~eV$, resulting 
in approximately $10^5$ simulated particles.

The complex plasma-chemistry of oxygen gives rise to a large number of 
non-reactive and reactive collisions~\cite{KG03}. In table~\ref{reactions} 
we show the ones with the largest cross sections. They are included in our model
and will be discussed in 
detail in Section III. We treat only three species  kinetically: electrons ($e$), 
negatively charged oxygen atoms ($O^-$), and positively charged oxygen molecules 
($O_{2}^{+}$). Neutral particles appearing either as educts or products in 
table~\ref{reactions} are not explicitly simulated. Molecular oxygen in its 
ground state, $O_2$ (feedstock gas), is modelled as a reservoir
characterized by a pressure $p$ and a temperature $T$. Whereas atomic oxygen, $O$, 
vibrationally excited oxygen molecules, $O_2(\nu)$, the Rydberg state of the 
molecular oxygen, $O_2({\rm Ryd})$, and the molecular meta-stables, $O_2(a^1\Delta_g)$ 
and $O_2(a^1\Sigma_g)$, are only indirectly accounted for in as 
far as their production results in an energy loss for electrons. 

The meta-stable $O_2(a^1\Delta_g)$ requires special attention because 
it also appears as an educt in the entrance channel for associative detachment (17).
Its concentration is therefore important, and we should actually build-up 
the $O_2(a^1\Delta_g)$ distribution function, that is, we should also 
simulate $O_2(a^1\Delta_g)$ particles. In that case, however, not only their 
volume production process (11) but also their volume and surface loss processes 
should be included. 
This is beyond the one-dimensional model. To take associative 
detachment, which is known to be an important 
process~\cite{SNM96,KSQ00,Franklin01a,BEL05}, nevertheless into
account, we use instead a simple model with one free parameter, which can
be interpreted as the $O_2(a^1\Delta_g)$ to $O_2$ density ratio (see below). 
 


\begin{figure}[t]
\begin{center}
\includegraphics[width=0.90\linewidth]{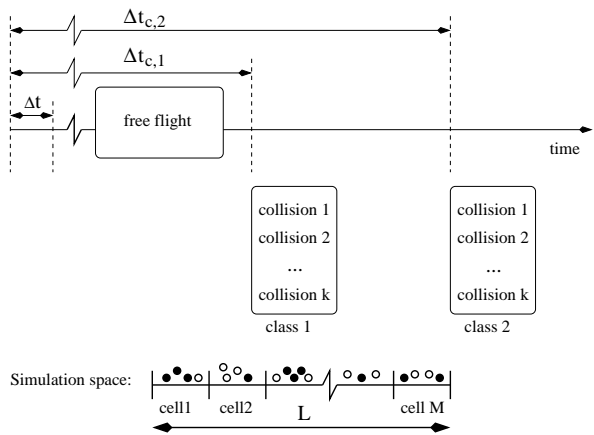}
\caption{\label{Alg}
Graphical representation of our implementation of the PIC-MCC
scheme.
}
\end{center}
\end{figure}

The three-species PIC-MCC model describes the physics of an ${\rm O}_2$ discharge 
by tracing the spatio-temporal evolution of an ensemble of charged 
particles ($e$, ${\rm O}^-$, and ${\rm O}_2^+$). As usual, to track the velocities 
and positions of the simulated particles, the simulation domain $0\le x \le L$ and 
the time are discretized and free flights are decoupled from  
collisions (see figure~\ref{Alg}). 
In order to have stable free flights of charged particles (in the mean electric
field given by the Poisson equation), the spatial resolution $\Delta x$ has to be 
less then the electronic Debye length, $\lambda_{De}$, and the 
time step $\Delta t$ should resolve the electron plasma oscillation whose frequency 
is $\omega_{pe}$~\cite{Birdsall85,Hockney89}. Both $\lambda_{De}$ and $\omega_{pe}$ are
set by the RES. 

The time interval over which collisions and free flights are decoupled is of the 
order of the mean free time, which, in general, is much larger then the fundamental  
time step $\Delta t$ enforced by the electric field. We distinguish two groups of 
collisions, each of which characterized by an average mean free time. The first 
group of collisions, occurring on a time scale $\Delta t_{c,1}$ comprises all 
collisions of table~\ref{reactions} except Coulomb collisions between charged particles, 
which take place on a different time scale $\Delta t_{c,2}$.  

To build up a representative sample for the charged particles, we 
first sample the initial distribution functions for $e$, ${\rm O}^-$, and ${\rm O}_2^+$ 
and then repeat the following procedure~\cite{Matyash03} (see figure~\ref{Alg}): 
\begin{itemize}
\item[(i)] For $n_{1}=\Delta t_{c,1}/\Delta t$ time steps, the coordinates 
    and velocities of all simulated particles are moved according to Newton's 
    equations of motion. The force acting on particles with charge $q_i$ 
    is given by ${\bf F_i}=-q_i\nabla\Phi$ where $\Phi$ is the electric potential 
    satisfying the Poisson equation with the boundary conditions specified above.
    Force interpolation and charge assignment are done within a linear
    interpolation scheme.
\item[(ii)] After $n_{1}$ time steps, a representative sample of the first class
     of collisions is executed, with post-collision velocities stochastically 
     determined in accordance with energy and momentum conservation, followed by  
     free flights with $n_1-n_2$ time steps, where $n_2=\Delta t_{c,2}/\Delta t$, 
     after which a representative random sample of Coulomb collisions 
     is performed, again consistent with energy and momentum conservation.  
\end{itemize}
After typically $10^3$ rf cycles quasi-stationarity with respect to the 
rf cycle is achieved, that is, within the statistical noise, macroscopic 
quantities, for instance, density and potential profiles, do not change anymore when 
averaged over a rf cycle.

Our treatment of collisions utilizes a number of concepts originally developed for 
the direct simulation Monte Carlo of rarefied gas flows~\cite{Bird69,Bird94}. Since 
it differs from the handling of collisions in the standard PIC-MCC schemes, which 
is based on the null-collision method~\cite{VS95}, we give some more details. 
After a free flight is completed, we sort the simulated particles into cells 
(see figure~\ref{Alg}). Simulated and feedstock gas particles located in one and 
the same cell (collisions are local) have then the opportunity to sequentially perform 
all possible types of collisions with each other. Simulated particles produced in the 
exit channel of a collision are readmitted for the next collision in the list. The ordering 
of particles into cells has to be repeated after each collision producing or annihilating 
simulated particles. Thus, as in the real system, simulated particles may perform more 
than one type of collision within the time interval over which the simulation decouples 
free flights and collisions. 

The probability of a simulated particle $i$ to make a collision of type $j$ is 
given by
\begin{eqnarray}
P_j^{(i)}=n_j\sigma_j u_j^{(i)} \tau_j~.
\end{eqnarray}
It depends on the density $n_j$ of the collision partners which 
act as the targets, the total cross section $\sigma_j$ for the collision, 
the relative velocity $u_j^{(i)}$ between particle $i$ and a particular 
collision partner, and the mean free time $\tau_j$. As can be seen 
in table~\ref{reactions}, except for associative detachment (17), which we 
discuss in more detail in the next section, the three species model recruits 
collision partners either from the simulated particles or from the feedstock 
gas ${\rm O_2}$. In the first case, $n_j$ is the number of simulated particles 
of the respective type in the considered cell divided by the cell volume, 
whereas in the second case, $n_j=p/kT$,
with $p$ the gas pressure and $T$ the room temperature. We do not calculate the 
mean free times $\tau_j$ for all the collisions listed in table~\ref{reactions}. 
Instead, we set $\tau_j=\Delta t_{c,2}$ for Coulomb collisions (1)~--~(3) and 
$\tau_j=\Delta t_{c,1}$ for the remaining collisions (4)~--~(19).


In order to verify that our code treats collisions correctly, we performed 
various test runs and compared the results either with analytical results~\cite{Matyash03} 
or with results obtained from the BIT1 code~\cite{TK02}. The latter was also 
used to check discrepancies in the cross section data. 

\section{III. Scattering and reaction channels}

\begin{table}
\caption{\label{reactions} Collisions included in our model}
\begin{ruledtabular}
\begin{tabular}{ll}
elastic scattering                                      & \\
(1) ${e + e \rightarrow e + e}$                         & \\
(2) ${\rm O^- + O^- \rightarrow O^- + O^-}$             & \\
(3) ${\rm O_2^+ + O_2^+ \rightarrow O_2^+ + O_2^+}$     & \\
(4) ${\rm e + O_2 \rightarrow e + O_2}$             & \\
(5) ${\rm O^- + O_2 \rightarrow O^- + O_2}$             & \\
(6) ${\rm O_2^+ + O_2 \rightarrow O_2 + O_2^+}$         & charge exchange\\
electron energy loss scattering                         & \\
(7) ${\rm e + O_2 \rightarrow e + O_2(\nu=1,...,4)}$    & vibrational excitation\\
(8) ${\rm e + O_2 \rightarrow e + O_2(Ryd)}$            & Rydberg excitation  \\
(9) ${\rm e + O_2 \rightarrow e + O(3P) + O(3P)}$       & dissociation (6.4~eV)\\
(10) ${\rm e + O_2 \rightarrow e + O(3P) + O(1D)}$       & dissociation (8.6~eV)\\
(11) ${\rm e + O_2 \rightarrow e + O_2(a^1\Delta_g)}$   & meta-stable excitation\\
(12) ${\rm e + O_2 \rightarrow e + O_2(b^1\Sigma_g)}$   & meta-stable excitation\\
electron $\&$ ion production $\&$ loss                  & \\
(13) ${\rm e + O_2^+ \rightarrow O + O}$                & dissociative recombination\\
(14) ${\rm O^- + O_2^+ \rightarrow O + O_2}$            & neutralization\\
(15) ${\rm e + O_2 \rightarrow O + O^-}$                & dissociative attachment\\
(16) ${\rm O^- + O_2 \rightarrow O + O_2 + e}$          & direct detachment\\
(17) ${\rm O^- + O_2(a^1\Delta_g) \rightarrow O_3 + e}$ & associative detachment\\
(18) ${\rm e + O_2 \rightarrow 2e + O_2^+}$             & impact ionization\\
(19) ${\rm e + O^- \rightarrow O + 2e}$                 & impact detachment\\
\end{tabular}
\end{ruledtabular}
\end{table}
Collisions strongly influence the particle concentration and the 
energy balance in the discharge. For oxygen, an
overwhelming number of elastic, inelastic, and reactive collisions is possible. 
Restricting, however, the description to a three species plasma ($e$, ${\rm O}^-$, 
and ${\rm O}_2^+$) many processes can be ignored and others can be treated approximately. 
In table~\ref{reactions} we show the collisions defining our three species model for an 
oxygen discharge, and classify them into three groups: elastic scattering, 
electron energy loss scattering, and electron and ion production
and loss reactions. The respective cross sections 
for these processes are shown in figures~\ref{ElasticXs}~--~\ref{ProdLossXs} as a 
function of the relative energy of the two particles in the entrance channel of 
the respective process. 

Our collection of cross sections is semi-empirical, combining measured data, 
which is usually available only in a finite energy range, with simple models for the 
low-energy asymptotic, which in most cases is not very well known from experiments. 
In general, the high-energy asymptotic has to be also determined from models, 
but it is less critical because the distribution functions usually decay sufficiently
fast at high energies. If not stated otherwise, we extrapolated therefore the values 
of the cross sections for the largest energies shown in the plots to all energies 
above it. Some of the cross sections significantly deviate from the ones previously
used~\cite{VS95}. Our simulations indicate, however, that the modifications 
are essential for obtaining bulk densities in accordance with 
experiments~\cite{KSQ00,Dittmann07}. 

\subsection{A. Elastic scattering}

Elastic collisions (1)~--~(6) are particle number conserving. Thus, when a
collision takes place, only the post-collision velocities of the scattering
partners have to be determined whereas the list of simulated particles
remains unaltered. 

The first three scattering processes (1)~--~(3) are intra-species Coulomb collisions.
They are not very important for the experiments~\cite{KSQ00,Dittmann07} we analyze 
in this series of papers. For other parameter regimes, however, the {\it local} ion density 
in an electro-negative gas discharge can be rather high~\cite{Kaganovich95} and it cannot be 
ruled out that ion-ion Coulomb collisions affect, for instance, the width of these ion 
density peaks. Since the computational burden is moderate, we kept intra-species Coulomb 
scattering in all our simulations. If not stated otherwise, inter-species Coulomb 
collisions were however neglected.

For Coulomb collisions, we used a binary collision model~\cite{TA77} 
with a uniformly distributed azimuth angle $\phi$, and a 
Gaussian distribution for $\tan\chi/2$, where $\chi$ is the scattering angle.
The second moment of the Gaussian distribution is given by~\cite{TA77}
\begin{eqnarray}
\langle\tan^2{\frac{\chi}{2}}\rangle=\frac{q^4n\ln\Gamma}
                              {8\pi\epsilon_0 m^2u^3}t_{c,2}~,
\end{eqnarray}
with $t_{c,2}$ the collision time for Coulomb scattering, 
$\ln\Gamma$ the Coulomb logarithm, $u$ the magnitude of the relative 
velocity in the center-of-mass frame, and $n$, $q$, $m$, and $\epsilon_0$
the local density, the charge, the reduced mass of the charged species 
under consideration, and the dielectric constant of the vacuum, respectively. 
\begin{figure}[t]
\includegraphics[width=0.90\linewidth]{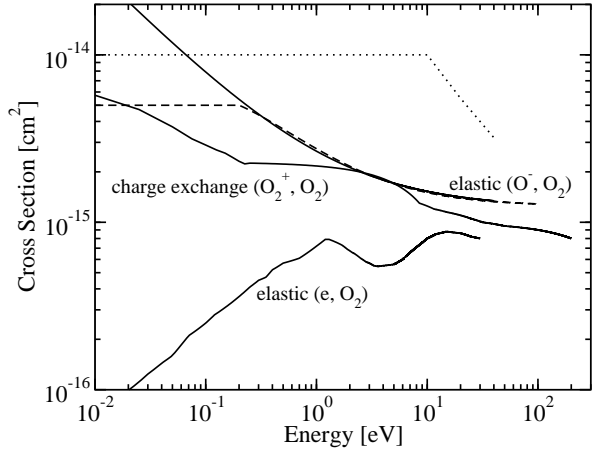}
\caption{\label{ElasticXs}
Cross sections for elastic scattering of electrons (4), ${\rm O}^-$ ions (5),
and ${\rm O}_2^+$ ions (6) on oxygen molecules. The dashed line and the dotted 
line indicate, respectively, the cross section for elastic (${\rm O}^-, {\rm O}_2$) 
scattering and the cross section for (${\rm O}^+_2, {\rm O}_2$) charge exchange 
scattering used in  Ref.~\cite{VS95}.
}
\end{figure}

Figure~\ref{ElasticXs} shows the cross sections for elastic scattering of 
electrons (4) and ${\rm O}^-$ ions (5) on ${\rm O}_2$ molecules  
together with the cross section for charge exchange scattering (6) of 
${\rm O}_2^+$ ions on ${\rm O}_2$ molecules. Since little is known about the
angle dependence of these processes, we assume the collisions to 
be isotropic, with a uniform distribution of the azimuth angle $\phi$ and a 
uniform distribution of $\cos{\chi}$, where $\chi$ is again the scattering
angle.

The cross section for elastic scattering of electrons on
${\rm O}_2$~\cite{Phelps85} is the same as in Ref.~\cite{VS95}. The other 
two cross sections are different. 
For charge exchange scattering, we constructed the cross section as follows. 
Below $0.251~eV$ and above $8.5~eV$, we used 
empirical data for momentum transfer scattering~\cite{STS63,GR72}, together 
with the expression
\begin{eqnarray}
\sigma_{cx}(E)=\frac{1}{2}\sigma_m(E)~,
\end{eqnarray}
where $\sigma_m$ and $\sigma_{cx}$ denote, respectively, the momentum and
charge exchange cross section.  For energies in between, we employed a 
linear interpolation. With this cross section, we could reproduce experimentally 
measured ${\rm O}_2^+$ velocity 
distribution functions very well~\cite{Matyash07}. With the charge exchange  
cross section given in Ref.~\cite{VS95} (dotted line in figure~\ref{ElasticXs}), 
on the other hand, we could not obtain the 
correct distribution functions -- neither with our PIC-MCC code nor with the BIT1 
code~\cite{TK02}, which we used in addition to exclude possible mistakes in our 
collision routines. 

The cross section for elastic scattering of ${\rm O}^-$ on ${\rm O}_2$ originates 
from a model which assumes polarization scattering between the two, induced by a
central potential $V(r)=-C/r^n$ with $n=3$. It is given by
\begin{eqnarray}
\sigma_{e}(E)=
\big[\frac{n\pi}{n-2}\big(\frac{n-2}{2}\cdot\frac{C}{E[eV]}\big)^{2/n}+\sigma_0\big]
\cdot cm^2~,
\label{Okada}
\end{eqnarray}
with $C=3.77\cdot 10^{-24}$ and $\sigma_0=1.2\cdot 10^{-15}$~\cite{Muschlitz60,OST78}. 
Here, we added a constant shift $\sigma_0$ to match the cross section at high energies as 
given in Ref.~\cite{VS95}. 

\subsection{B. Electron energy loss scattering}
\begin{figure}[t]
\includegraphics[width=0.90\linewidth]{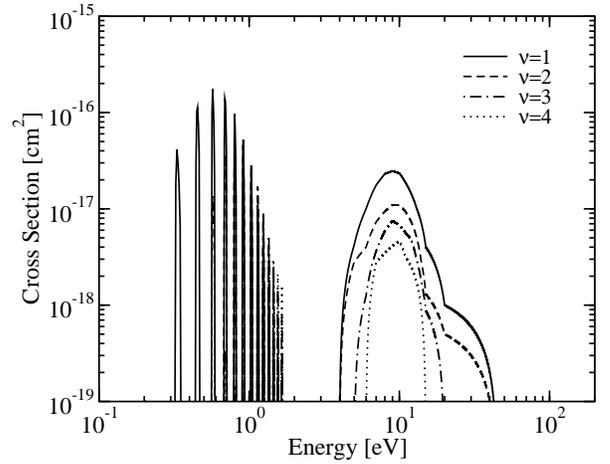}
\caption{\label{InelasticXs_v}
Cross sections for the $\nu=1,2,3,$ and $4$ vibrational excitation of
${\rm O}_2$ molecules (7).
}
\end{figure}

Electron energy loss occurs due to inelastic collisions, (7)~--~(12), in which
the oxygen molecule is either promoted into an excited state or dissociated into 
neutral fragments. In the three species model, where the spatio-temporal 
evolution of the neutral fragments and the excited oxygen molecules is not traced, 
we can treat this group of processes in the spirit of ``test particle collisions'' 
where the electron is the ``test particle'', suffering momentum as well as energy 
transfer, and the ${\rm O}_2$ molecule is the ``field particle'' (with internal 
degrees of freedom). Drawing the field particles from a Maxwell 
distribution characterized by the gas temperature $T$ and the gas density $n=kT/p$,
the same binary collision model can be used as for elastic scattering~\cite{TA77}. 
The only difference is that now the magnitude of the 
post-collision relative velocity in the center-of-mass system is given by 
\begin{eqnarray}
u'=\sqrt{u^2-\frac{2\delta E}{m_{\alpha\beta}}}~,
\end{eqnarray}
where $u$ is the magnitude of the pre-collision relative velocity in the 
center-of-mass frame, $m_{\alpha\beta}$ is the reduced mass of the scattering 
partners, which, in general, have not the same mass, and 
$\delta E$ is the excitation energy of the process. Due to lack of 
angle-resolved scattering cross sections, we again assume the collisions to be 
isotropic.
\begin{figure}[t]
\includegraphics[width=0.90\linewidth]{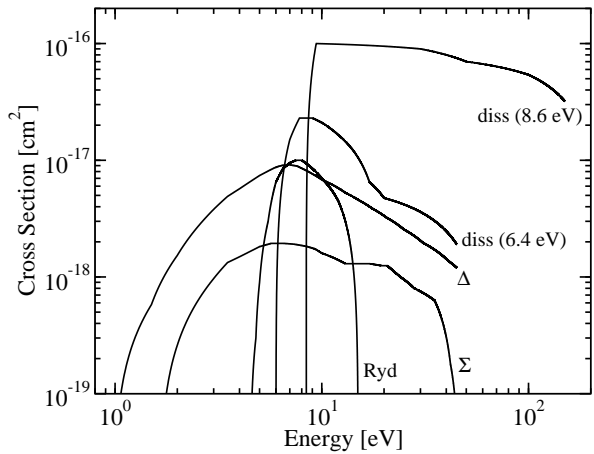}
\caption{\label{InelasticXs_e}
Cross sections for Rydberg excitation (8), $6.4~eV$ (9) and $8.6~eV$ (10)
dissociative excitation, $\Delta$ excitation (11), and
$\Sigma$ excitation (12) of ${\rm O}_2$ molecules.
}
\end{figure}

The most important electron energy losses are due to vibrational and electronic 
excitation and impact dissociation of the oxygen molecule. The cross sections for 
these processes~\cite{Phelps85} are the same as the ones used in Ref.~\cite{VS95}.
For completeness, they are plotted in figures~\ref{InelasticXs_v} and 
\ref{InelasticXs_e}. We consider vibrational excitations with 
$\nu=1,2,3$, and $4$, excitation of the ${\rm O}_2$ Rydberg state, $PP$ and $PD$ 
dissociation, and the excitation of the meta-stable states ${\rm O_2(a^1\Delta_g})$ 
and ${\rm O_2(b^1\Sigma_g})$. In contrast to Ref.~\cite{VS95}, we ignore
$DD$ dissociation, because, in the energy range of interest, its cross section 
is one order of magnitude smaller than for $PP$ and $PD$ dissociation, and 
rotational excitation which is almost elastic (small energy transfer)
and thus dominated by the much more efficient elastic scattering of electrons
on ${\rm O}_2$. 

\subsection{C. Electron and ion production and loss reactions}

Electron and ion production and loss reactions (13)~--~(19) are inelastic
collisions which change the number of charged particles. Therefore, the
list of simulated particles has to be up-dated. All collisions obey 
conservation laws and are again assumed to be isotropic. The details of 
the modeling depend on the process.  

The two recombination channels (13) and (14) result simply in the annihilation
of the two oppositely charged particles which participate in the process. 
Binary collisions with one charged particle in the exit channel, such as  
dissociative attachment (15) and associative detachment (17), are treated
within the modified binary collision model described in the previous
subsection. Impact ionization (18) and impact detachment (19) are modelled
as follows: First, an inelastic binary collision is performed, in which 
the parent electron looses the ionization (detachment) energy. Then, the 
post-collision ${\rm O}_2$ (${\rm O}^-$) particle is split into an electron
and a ${\rm O}_2^+$ (${\rm O}$) particle. Finally, an elastic binary 
collision is applied to distribute energy among the two charged 
particles. Modeling three particle processes as a sequence of two binary 
collisions with a particle splitting in between guarantees energy 
and momentum conservation which is critical for the stability of 
simulations~\cite{Matyash03}. Although direct detachment (16) 
could be modelled in the same spirit, we adopted a simpler approach. From 
experiments~\cite{BM70} we know that the energy of the released electron 
is approximately one-tenth of the energy of the primary ${\rm O^-}$ ion.
To obtain the energy distribution of the ejected electron we hardwired 
therefore this ratio in the collision routine for direct detachment. 

\begin{figure}[t]
\includegraphics[width=0.90\linewidth]{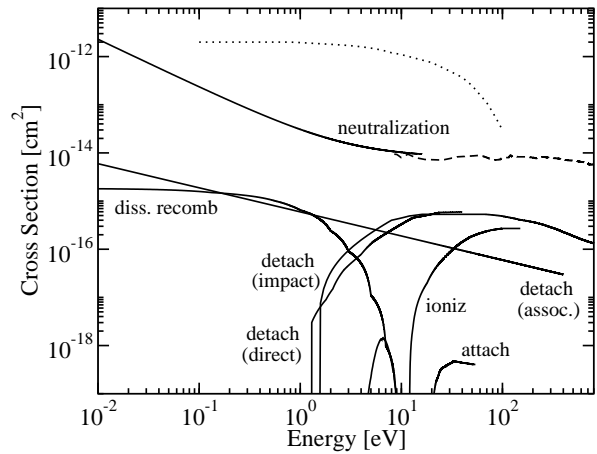}
\caption{\label{ProdLossXs}
Cross sections for dissociative recombination (13),
neutralization (14), dissociative attachment (15), direct detachment (16),
associative detachment (17), impact ionization (18), and impact detachment (19). 
The dashed line indicates the experimentally determined high-energy asymptotic of
the neutralization cross section and the dotted line is the 
cross section for neutralization used in Ref.~\cite{VS95}.
}
\end{figure}

The cross sections for electron and ion production and loss reactions are shown 
in figure~\ref{ProdLossXs}. For impact ionization~\cite{Phelps85}, dissociative 
attachment~\cite{RB65}, and impact detachment~\cite{TB68,PFD79,VKM96} we used 
empirical cross sections throughout, putting, in the case of impact detachment 
a higher confidence level to more recent data~\cite{VKM96}, whereas for 
dissociative recombination, ion-ion neutralization, and detachment on neutrals 
we combined empirical data with analytical models.  

First, we discuss dissociative recombination (13). For this process we used an 
effective cross section, 
\begin{eqnarray}
\sigma_{\rm dr}(E) = B \cdot \sigma_0(E)~,
\end{eqnarray}
where $\sigma_0$ is the cross section for non-resonant dissociative 
recombination~\cite{Bardsley68} and $B=3$ is a scaling factor, which is used 
to fit experimental data~\cite{WD74}. A comparison of the effective cross section
with a calculated cross section~\cite{AKO82} shows that $B$ can be 
also considered as a simple device to approximately take the effect of resonant
dissociative recombination channels into account which originate from 
vibrationally excited states of $O_2^+$~\cite{AKO82}.

For ion-ion neutralization (14), we constructed a cross section from a two-channel 
Landau-Zener model~\cite{Olson72}, with one free parameter, which we adjusted to obtain
the correct high-energy asymptotic of the cross section~\cite{PP98}. Ion-ion neutralization 
occurs because the adiabatic energy of the $({\rm O^-},{\rm O}_2^+)$ configuration 
decreases when the ions approach each other. At a certain distance $R_x$, the energy 
of the $({\rm O^-},{\rm O}_2^+)$ configuration falls below the energy of 
the $({\rm O},{\rm O}_2)$ configuration~\cite{Lieberman05}. The Landau-Zener theory 
estimates the probability for switching from one configuration to the other at the 
distance $R_x$ and leads to a cross section,
\begin{eqnarray}
\sigma_{r}(E) = 4\pi R_x^2\big(1+\frac{1}{R_x E}\big)~,
\end{eqnarray}
which approaches a constant $C=4\pi R_x^2$ for $E R_x\gg 1$. Fitting $C$ and thus 
$R_x$ to empirical data at high energies~\cite{PP98}, we obtain
\begin{eqnarray}
\sigma_r(E) = 0.8\cdot\big(1+\frac{2.85}{E[eV]}\big)\cdot 10^{-14}\cdot cm^2~.  
\label{Xsection_recomb}
\end{eqnarray}
This cross section deviates dramatically from the one used in Ref.~\cite{VS95}
(dotted line in figure~\ref{ProdLossXs}).

The most severe modification we made was for detachment of ${\rm O}^-$ on neutrals.
It takes place through direct detachment (16), resulting in an atomic 
oxygen, an oxygen molecule and an electron, and through associative detachment 
(17), which leads to an ${\rm O}_3$ molecule and an electron. The latter is rather 
surprising because there is no evidence for it in beam experiments (where only 
direct detachment is observed~\cite{CS74a,CS74b}). Yet, experimental studies of
${\rm O}_2$ discharges~\cite{KSQ00,BEL05} (as well as general theoretical
considerations~\cite{Franklin02,SNM96,Franklin01a}) 
strongly suggest that associative detachment is possible in an oxygen discharge 
because of the presence of meta-stable ${\rm O}_2(a^1\Delta_g)$. In contrast to 
direct detachment, which has a threshold around $1.3~eV$~\cite{CS74a} 
(see figure~\ref{ProdLossXs}), associative detachment has usually no threshold. 
Thus, it may be a rather important loss channel for cold ${\rm O}^-$ ions.

Since we could not find an empirical cross section for associative detachment, we employed 
a model, which describes the detachment (electron loss) as the ``inverse'' of a classical 
Langevin-type electron capture into an attractive auto-detaching state of ${\rm O_3^-}$. 
Assuming the polarizability for ${\rm O}_2(a^1\Delta_g)$ 
to be the same as for ${\rm O}_2$, the cross section is then given by~\cite{Lieberman05}
\begin{eqnarray}
\sigma^\Delta_{ad}(E) =
5.96 \cdot \frac{10^{-16}\cdot cm^2}{\sqrt{E[eV]}}~.
\label{Xsection_ad}
\end{eqnarray}
As can be seen in figure~\ref{ProdLossXs}, associate detachment is already the dominant 
detachment process for energies below $\approx 6~eV$. 
\begin{figure}[t]
\includegraphics[width=0.90\linewidth]{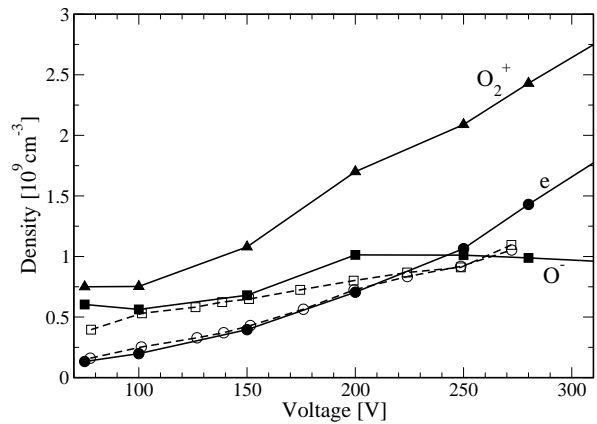}
\caption{\label{Central}
Voltage dependence of the quasi-stationary, cycle-averaged electron and ion densities at 
$x=L/2$ (bulk densities) for a discharge
with $L=2.5~cm$, $p=13.8~Pa$, and $f_{rf}=13.6~MHz$. Filled and open symbols denote,
respectively, results of the simulation and experimentally measured 
densities~\cite{KSQ00}. Solid and dashed lines guide the eye.
}
\end{figure}

From the cross section alone, of course, we cannot determine the probability $P_{ad}$ 
for associative detachment. We also need the density of ${\rm O_2(a^1\Delta_g)}$,
which is unknown in the three species model. However, it should be of the order
of the ${\rm O}_2$ density. Therefore, we write $n_\Delta = C\cdot n_{{\rm O}_2}$, 
with $C<1$, and obtain $P_{ad}=u\cdot\sigma_{ad}\cdot (C \cdot n_{{\rm O}_2})\cdot t_{c,1}$,
where $C$ is a fit parameter which can be adjusted to experiments. 

\section{IV. Comparison with experiment}

The plasma model described in the previous sections contains three parameters: 
$\sigma_0$, $B$, and $C$. The first two we fixed by a direct comparison of the 
model cross section with the measured cross section. The parameter $C$, 
in contrast, cannot be determined in this way, because it is not linked to a cross 
section. It denotes instead the fraction of ${\rm O}_2$ molecules in the meta-stable  
${\rm O_2(a^1\Delta_g)}$ state which in turn depends on the particular set-up of the 
discharge. We consider therefore $C$ as a free parameter which 
can be adjusted to the discharge to be modelled.

To validate our model we 
focus now on the oxygen discharge described in Ref.~\cite{KSQ00}. In order to determine 
$C$, we simulated the discharge for $p=13.8~Pa$, $U_{rf}=250~V$, and $f_{rf}=13.6~MHz$, 
and tuned $C$ to reproduce the quasi-stationary, cycle-averaged negative and positive 
ion densities at $x=L/2$ (bulk densities); the bulk electron density matches then 
also because of quasi-neutrality in the bulk of the discharge. We obtained $C\approx 1/6$, 
implying that roughly one sixth of the ${\rm O}_2$ molecules is in the meta-stable 
state. 

In figure~\ref{Central} we plot the bulk ion and electron densities for 
$p=13.8~Pa$ over a wide voltage range, using however for all voltages $C\approx 1/6$, 
the value determined for $U_{rf}=250~V$. As can be seen, the agreement between simulation
and experimental data~\cite{KSQ00} is rather good, indicating that our model captures 
the essential processes in the bulk of an oxygen discharge.

\begin{figure}[t]
\includegraphics[width=0.90\linewidth]{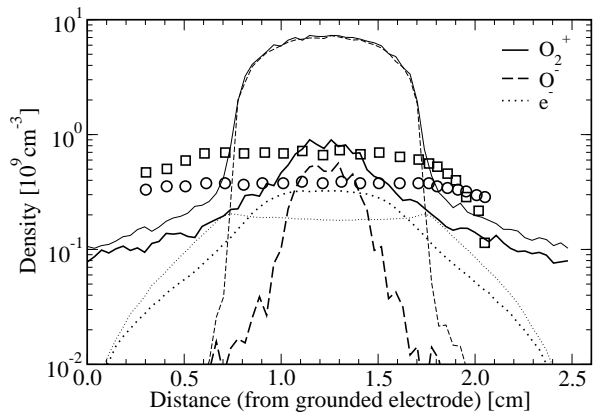}
\caption{\label{Profiles}
Quasi-stationary, cycle-averaged electron and ion density profiles for $U=150~V$
without (thin lines) and with (thick lines) associative detachment taken into account;
$p$, $L$, and $f_{rf}$ are the same as in figure~\ref{Central}. Open circles and open
squares denote, respectively, the measured electron and ${\rm O}^-$ density
profiles~\cite{KSQ00}. Note, the runs producing the shown data included
inter-species Coulomb scattering.
}
\end{figure}

The precise value of $C$ should not be taken too serious because it is
based on a rather crude model for associative detachment.
More important is that without this process ($C=0$), the simulation
could not reproduce the measured densities.
This can be seen in figure~\ref{Profiles}, where we plot 
the quasi-stationary, cycle-averaged density profiles without (thin lines) and with 
associative detachment (thick lines) taken into account. (For these runs, we also 
included inter-species Coulomb collisions but the profiles without it are basically
the same.) Neglecting associative detachment, 
the bulk ion densities turned out to be almost one order of magnitude too high.  
Remarkably, without associative detachment, the density profiles have the parabolic 
shape expected from ambipolar drift-diffusion 
models~\cite{EE80,FGT88,Tsendin89,LVL94,FS99,FS00,LMF04}. 

\begin{figure}[t]
\includegraphics[width=0.90\linewidth]{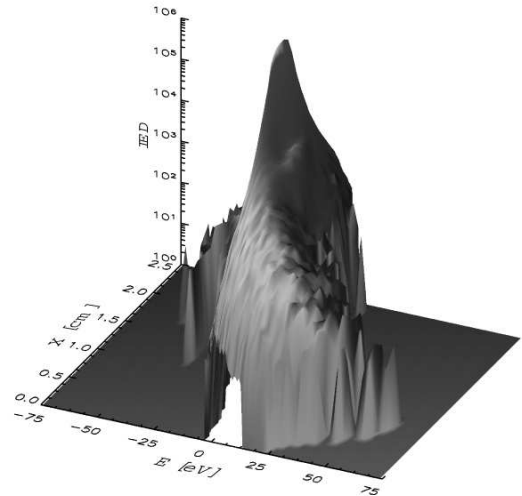}
\caption{\label{fminus}
Quasi-stationary, cycle-averaged negative ion velocity distribution function 
$f_-(E,x)$ with $E={\rm sign}(v_x)v_x^2/2M_-$ for $U_{rf}=150~V$; $p$, $L$, 
and $f_{rf}$ as in figure~\ref{Central}.
}
\end{figure}

To understand why associative detachment is so crucial for modeling the discharge
of Ref.~\cite{KSQ00}, we plot in figure~\ref{fminus}
for $p=13.8~Pa$ and $U_{rf}=150~V$ the (marginal) velocity distribution function 
for negative ions, $f_-(E,x)$ with $E={\rm sign}(v_x)v_x^2/2M_-$. Clearly, the majority 
of ${\rm O}^-$ ions 
is cold; the distribution function spreads out only in the region where the bulk plasma merges
with the peripheral plasma at $x\approx 0.8~cm$. Thus, loss processes whose cross sections 
are large for small energies (see figure~\ref{ProdLossXs}), that is, ion-ion neutralization 
and associative detachment, will be very efficient. Direct detachment, on the other 
hand, whose cross section has a threshold, will be suppressed. At low 
energies, both ion-ion neutralization and associative detachment have cross sections 
which increase with decreasing energy. Which process dominates depends therefore also 
on the collision probability $P$. For ion-ion neutralization $P$ is proportional to  
the ${\rm O}_2^+$ density while for associative detachment it is proportional to 
the ${\rm O_2(a^1\Delta_g)}$ density. Since for the discharge of Ref.~\cite{KSQ00}, 
the ${\rm O}_2^+$ density is much smaller than the ${\rm O_2(a^1\Delta_g)}$ density, 
which we estimated to be $1/6$ of the ${\rm O}_2$ density, associative detachment 
has to be the main loss process for negative ions in this experiment.

\begin{figure}[t]
\includegraphics[width=0.90\linewidth]{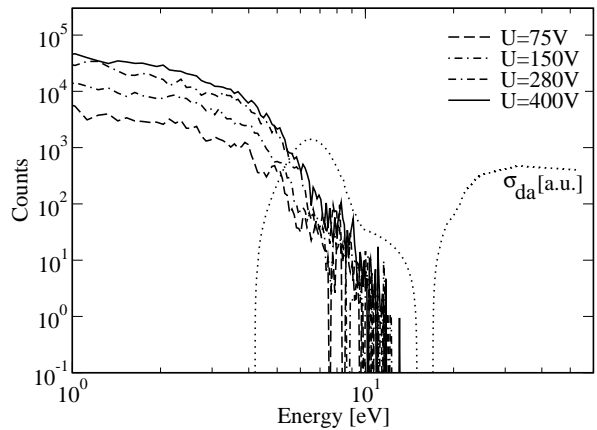}
\caption{\label{felectron}
Quasi-stationary, cycle-averaged electron velocity distribution function $f_e(E,x)$
with $E=v^2/2m_e$ and $x=L/2$ for $U_{rf}=75~V,
150~V, 280~V$, and $300~V$; $p$, $L$, and $f_{rf}$ as in figure~\ref{Central}. The
dotted line indicates the cross section for dissociative attachment (15) in 
arbitrary units.
}
\end{figure}

Additional support for this claim comes from the weak voltage dependence of the 
${\rm O}^-$ density between $75~V$ and $300~V$ indicating that both ${\rm O}^-$ 
production and ${\rm O}^-$ loss are in this voltage range rather insensitive to 
voltage changes. We will now show that this strongly suggests that associative 
detachment is the main loss process for ${\rm O}^-$ ions. First, we conclude 
from figure~\ref{felectron}, which shows the attachment cross section 
(in arbitrary units) together with bulk electron velocity distribution 
functions for different voltages, that the energy range
between $4~eV$ and $15~eV$ will be the most important one for the attachment
process; for energies below $4~eV$ the production is zero because 
the cross section is zero while for energies above $15~eV$ the number of 
electrons available for attachment is too low. We also see that in 
the voltage range of interest, the electron velocity distribution function
does not change much in this energy range. 
Thus, the production process is almost voltage-independent and the observed 
near constancy of the ${\rm O}^-$ density has to be taken as a signature of 
the loss process. As already mentioned, direct detachment can be ruled out.
Ion-ion neutralization, on the other hand, would lead for $E\ll 1.4~eV$, 
that is, in the relevant energy range, to an energy-resolved rate 
coefficient $K_r(E)\sim\sqrt{E}\cdot\sigma_r(E)\sim 1/\sqrt{E}$  
which strongly increases with decreasing energy. Thus, even small, 
voltage-induced changes of the energy of negative ions would suffice 
to lead to noticeable modifications of the loss rate. What remains is  
associative detachment. The energy-resolved rate coefficient for this process 
is $K_{ad}^\Delta(E)\sim\sqrt{E}\cdot \sigma_{ad}^\Delta(E)\sim const$.
At the same time, the distribution function for ${\rm O^-}$ ions does not 
change much between $75~V$ to $300~V$. Thus, the loss rate due to associative 
detachment is nearly independent of voltage. Together with the 
voltage independence of dissociative attachment, this leads to the 
observed constancy of the ${\rm O^-}$ density. 

Our simulation reproduces the absolute values of bulk densities fairly well. 
The (axial) density profiles of the simulation decay however too fast compared to the
experimental ones (open symbols in figure~\ref{Profiles}). We attribute this to the fact 
that the simulations assume a constant ${\rm O}_2(a^1\Delta_g)$ density, whereas, in 
reality, there should be a density profile, because of the interplay of volume and 
surface loss and generation processes for ${\rm O}_2(a^1\Delta_g)$ molecules. 
In particular surface losses, that is, the
decay of ${\rm O}_2(a^1\Delta_g)$ molecules when they hit the boundary of the 
discharge should play an important role because they lead to a depletion of 
${\rm O}_2(a^1\Delta_g)$ molecules in the vicinity of the electrodes and thus to a 
reduction of associative detachment. A detailed investigation 
of the ``drag'' the ${\rm O}_2(a^1\Delta_g)$ density is expected to exert on 
the ${\rm O^-}$ density is however beyond the scope of the present paper.

\section{V. Conclusions}

We presented a planar, one-dimensional PIC-MCC model for a capacitively coupled 
rf discharge in oxygen which is capable to quantitatively describe experiments. 
Its main features are: (i) Only 
electrons, ${\rm O}^-$ ions, and ${\rm O}_2^+$ ions are treated kinetically. 
(ii) Neutral particles are only incorporated in as far as
they affect the particle and energy balance of simulated charged particles. 
(iii) A direct simulation Monte Carlo model for collisions has been used, with 
two groups of collisions, which differ in the collision times. 

The elementary processes of our three species model are shown in 
table~\ref{reactions}. We took the processes with the largest cross sections 
into account. When possible, we implemented empirical 
cross sections. In some cases, however, we combined them with analytical model 
cross sections to obtain the correct low-energy asymptotic. In particular, we used
a polarization-type scattering model for elastic $({\rm O}^-, {\rm O}_2)$ scattering,
a Landau-Zener-type model for $({\rm O}^-, {\rm O}_2^+)$ neutralization, and 
an inverse classical, Langevin-type capturing model for  
${\rm O}_2(a^1\Delta_g)$-induced associative detachment. To describe associative 
detachment due to ${\rm O}_2(a^1\Delta_g)$ without explicitly calculating the 
${\rm O}_2(a^1\Delta_g)$ density, we furthermore introduced a parameter $C$, 
which should be fitted to the particular oxygen discharge under consideration
and could be interpreted as the fraction of ${\rm O}_2$ molecules in the 
meta-stable ${\rm O}_2(a^1\Delta_g)$ state. 

As a first application of our model, we simulated the discharge of 
Ref.~\cite{KSQ00}. After we adjusted the parameter $C$ to reproduce
the bulk ion densities for $p=13.8~Pa$ and $U_{rf}=250~V$, we also 
obtained the correct bulk densities for other voltages.
The parameter $C\approx 1/6$, that is, approximately one sixth of the oxygen 
molecules are in the meta-stable state. We also pointed out that the weak 
voltage dependence of the bulk densities for fixed pressure indicates
that ${\rm O}^-$ losses due to associative detachment dominate the ones due 
to ion-ion neutralization. 

Although we could reasonably well describe electron and ion densities in the
bulk, the (axial) ion density profiles of the simulation are too narrow 
compared to the experimental ones. Most probably this is because we assumed
a constant ${\rm O}_2(a^1\Delta_g)$ density. In reality, however, 
${\rm O}_2(a^1\Delta_g)$ molecules decay when they hit the boundary
of the discharge. The ${\rm O}_2(a^1\Delta_g)$ density should therefore
decrease in the vicinity of the electrodes and with it associative detachment. 
To take this effect into account requires however a model which not only
treats electrons, ${\rm O}^-$ ions, and ${\rm O}_2^+$ ions kinetically, but
also ${\rm O}_2(a^1\Delta_g)$ molecules. 

Our concern in the present paper, which is the first in a series of three,
was the calibration and validation of a three species  
model for capacitively coupled rf discharges in oxygen. For that purpose, we 
used published experimental data for the bulk of the discharge~\cite{KSQ00}. 
The following 
paper~\cite{Dittmann07} describes the results of an experimental investigation 
of the sheath region of such a discharge, focusing, among others, on the 
spatial and temporal evolution of the
$844~nm$ double emission layer in front of the powered electrode.
In the third paper~\cite{Matyash07}, finally, we will use the three 
species model to simulate the sheath region. Besides a comparison 
of simulated with measured distribution functions for positive ions, 
we will also present a model for the 
$844~nm$ double emission layer. The three species model, including its 
approximate treatment of associative detachment, is sufficient for that purpose, 
because negative ions are negligible in the region from which the emission 
originates. Negative ions are only needed for the overall charge balance 
of the discharge. Their density profile per se is not crucial for the 
explanation of the $844~nm$ double emission layer as long as it 
features a vanishingly small density in front of the electrode. This is 
already accomplished by the three species model.\\

\begin{acknowledgments}
Support from the SFB-TR 24 ``Complex Plasmas'' is greatly acknowledged. 
We thank B. Bruhn, H. Deutsch, K. Dittmann, and J. Meichsner for 
valuable discussions and critical reading of the manuscript. 
K. M. and R. S. acknowledge funding by the Initiative and 
Networking Fund of the Helmholtz Association and F. X. B. acknowledges 
special funding 0770/461.01 by the state Mecklenburg-Vorpommern.
\end{acknowledgments}

\end{document}